\documentclass[aps,prc,showpacs,superscriptaddress]{revtex4}
 \usepackage{graphics}
 \usepackage{amsmath}
 \usepackage{epsfig}
 \usepackage[english]{babel}

\begin{document}

\title{Electron Capture $\beta$-Decay of $^7$Be Encapsulated in C$_{60}$:\\
Origin of Increased Electron Density at the $^7$Be Nucleus.}
\thanks{Submitted to Phys.Rev.C}

\author{E.V.~Tkalya}
\email{tkalya@srd.sinp.msu.ru}

\affiliation{Skobeltsyn Institute of Nuclear Physics, Lomonosov
Moscow State University,\\
Leninskie gory, Moscow, Ru-119991, Russia}

\affiliation{National Institute for Theoretical Physics, Stellenbosch
Institute of Advanced Study, \\
Private Bag X1, Matieland 7602, South Africa}

\author{A.V.~Bibikov}

\affiliation{Skobeltsyn Institute of Nuclear Physics, Lomonosov
Moscow State University,\\
Leninskie gory, Moscow, Ru-119991, Russia}

\author{I.V.~Bodrenko}

\affiliation{Skobeltsyn Institute of Nuclear Physics, Lomonosov
Moscow State University,\\
Leninskie gory, Moscow, Ru-119991, Russia}

\date{\today}

\begin{abstract}

We offer a new theoretical interpretation for the effect of enhanced
electron density at $^7$Be nucleus encapsulated in fullerene
C$_{60}$. Our {\it{ab initio}} Hartree-Fock calculations show that
electron density at the $^7$Be nucleus in $^7$Be@C$_{60}$ increase
due to {\it{}attractive} effective potential well generated by the
fullerene. The 2$s$ state in the isolated Be atom turns into 3$s$
state in the joint potential. This new state has higher energy, and
slightly larger amplitude at the Be nucleus than the previous 2$s$
state. Moreover the 3$s$ wave function has additional {\it{}node}
appeared at the distance $r \simeq 5a_B$ from the center. The node
imitates repulsion between the Be electron and the fullerene wall,
because the electron has zero probability to occupy this region. Such
imitation of the repulsion by means of the node in attractive
potential has direct physical analogy in the theory of
$\alpha$-$\alpha$ and $N$-$N$ nuclear interactions.

\end{abstract}

\pacs{23.40.-s, 21.10.Tg, 36.40.Cg, 61.48.+c}

\maketitle

\section{Introduction}

In the $^7$Be $\beta$-decay via electron capture (EC) process, the
nucleus absorbs an electron from the atomic or molecular shell and is
transformed to $^7$Li in the reaction $p+e^-\rightarrow{}n+\nu_e$.
The decay rate is proportional to the electron density at the nucleus
and therefore depends on the chemical environment of the radioactive
isotope. $^7$Be has been used in investigations of the electron
capture decay rate in various chemical states since the first studies
by Segre \cite{Segre-47} and Daudel \cite{Daudel-47}. By now,
experimentalists have published the results of more than sixty
measurements relating to $K$- and $L$-shell electron capture by
$^7$Be in different chemical forms and media.

In 2005-2007, two teams studied the EC decay of $^7$Be inside the
fullerene C$_{60}$ \cite{Ohtsuki-04,Ray-06,Ohtsuki-07}. It was
found, that the half-life of $^7$Be in metallic beryllium measured
at room temperature exceeds the one in $^7$Be@C$_{60}$ at room
temperature by 0.83\% \cite{Ohtsuki-04}, and by 1.5\%
\cite{Ohtsuki-07} if the latter is measured at 5$^{\circ}K$. This
difference between the $^7$Be EC $\beta$-decay constants is the
largest among available experiments. A density functional theory
(DFT) based numerical calculations of the electron density at the
Be nucleus have been presented in \cite{Ohtsuki-07,Morisato-08}
along with the experimental data. It was found that, in accordance
with the experiment, the electron density at Be encapsulated in
the center of C$_{60}$  is larger than the one at the nuclei in
metallic beryllium. Qualitatively, in the metal, the electrons are
shared, decreasing their local density at the nuclei, while the
beryllium atom in C$_{60}$ remains intact. A more careful
analysis, however, has shown that the electron density at
$^7$Be@C$_{60}$  is larger by 0.17\% even in comparison with that
in an isolated beryllium atom. The authors explain this result as
a ``compression'' of the Be's 2$s$ orbital inside C$_{60}$. The
reason of the compression may be the ``repulsive interaction''
between Be and the C$_{60}$ cage according to \cite{Lu-02} .

\section{Examination within the framework of Hartree-Fock based methods}

Detailed understanding of the structure of atoms encapsulated in
fullerens is important, in particular, for developing the concept of
the fullerene as an isolating cage which ``does not affect'' the
trapped single atom and ``protects'' it from the outer environment.
In the present work, we study the Be-C$_{60}$ interaction and its
effect on the electron density at the Be nucleus within the framework
of Hartree-Fock (HF) based methods. Although our value of the
relative decay rate difference between metallic Be and Be@C$_{60}$ is
in qualitative agreement with the experimental one as well as with
the results of the previous theoretical studies, we suggest a new
theoretical interpretation of the physical nature of the enhanced
electron density at Be in C$_{60}$.

We started with the structural optimization of Be position inside
C$_{60}$. The fullerene's geometry was taken from experiment
\cite{Johnson-91} (the length of the long and the short bonds are
1.448 {\AA} and 1.404 {\AA}, respectively) and fixed during the
optimization, as the endohedral doping has, as expected, a small
effect \cite{Lu-02}. The total energy of Be@C$_{60}$ complex at every
trial configuration was calculated by the restricted (singlet spin
state) Hartree-Fock method, with the 6-31G$^{**}$ molecular basis set
\cite{Francl-82,BS} in a cartesian form. Besides, the electronic
correlations were taken into account within the second order
perturbation method (MP2). For the calculations, we used our original
program which employs the resolution of the identity (RI) method for
the electron-electron interaction integrals and allows to perform the
Hartree-Fock based calculations for large systems with moderate
computational resources (see \cite{Artemyev-05,Nikolaev-08} for
details). For both the Hatrtee-Fock and the MP2 variants, the
optimization results in the position of the Be atom at the center of
the fullerene are in full agreement with the previous DFT based
studies \cite{Ohtsuki-07,Lu-02}. In order to evaluate the interaction
energy defined as $\Delta E =
E_{\textrm{Be}@\textrm{C}_{60}}-(E_{\textrm{Be}} +
E_{\textrm{C}_{60}})$, we have performed additional calculations of
the total energies of Be atom, $E_{\textrm{Be}}$, and the fullerene,
$E_{\textrm{C}_{60}}$. Besides, in a separate calculation, the basis
set superposition error was taken into account by the counterpoise
(CP) method \cite{Boys-70}. The results are summarized in Table
\ref{tab:DE}.

\begin{table}[h]
\caption{The Be-C$_{60}$ interaction energy in eV (Be is at the
center of the fullerene).} \label{tab:DE}

\begin{tabular}{c|c|c}
\tableline
                    &    HF   & HF+MP2 \\
\hline
 CP corrected      &  0.91   & -0.41  \\
 uncorrected       &  1.00   & -0.63  \\
\tableline
\end{tabular}
\end{table}

We therefore conclude that Be atom's equilibrium position at the
center of the fullerene belongs to the attractive region of the
Van-der-Waals interaction, i.e. the Be-C$_{60}$ interaction is
attractive and the Be@C$_{60}$ complex is stable in the ground state
with respect to decay to Be and C$_{60}$. This result is in contrast
with the one of Lu et.al. \cite{Lu-02} who concluded a ``slightly
repulsive'' Be-C$_{60}$ interaction from a DFT calculations, and
obtained the value of +1.05 eV for the interaction energy. We apply
their speculations to our Hartree-Fock results, which also give the
repulsive interaction as the pure HF method may not account of the
dispersion energy.


\begin{figure}[h]
\includegraphics[angle=-90,width=10cm]{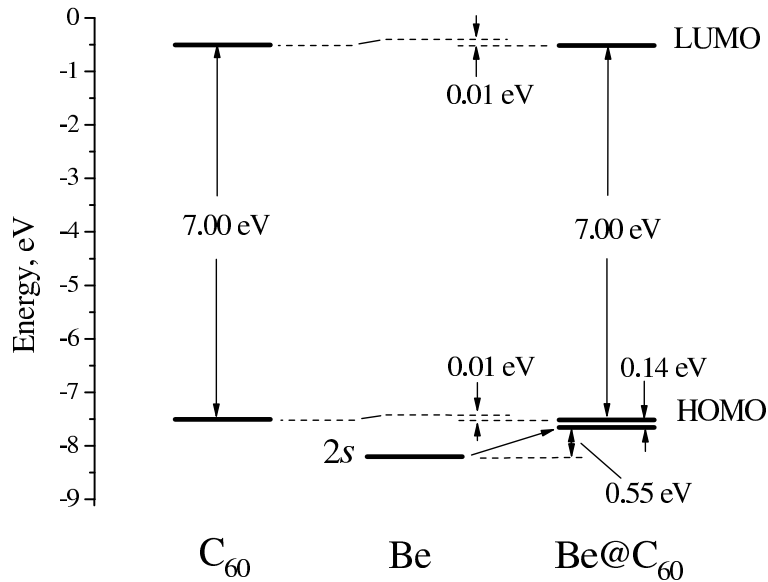}
\caption{Energy levels of molecular orbitals in C$_{60}$, Be atom and
Be@C$_{60}$.} \label{fig1}
\end{figure}

The consideration of Lu et.al. is based on the modifications in
the energy levels of the HOMO/LUMO (highest occupied / lowest
unoccupied molecular orbitals) of the Be atom and C$_{60}$ upon
formation of the Be@C$_{60}$ complex. Our results, presented in
Figure\ \ref{fig1}, differ qualitatively from those of
\cite{Lu-02}, calculated with the B3LYP density functional with
6-31G$^{**}$ molecular basis set. In \cite{Lu-02}, Be's atomic
2$s$ level lies in the middle of C$_{60}$'s LUMO-HOMO gap.
Moreover, C$_{60}$'s LUMO-HOMO gap increases in \cite{Lu-02} upon
endohedral doping with Be, because the HOMO is slightly lowered by
0.03 eV, while the LUMO is elevated by 0.04 eV. These results were
interpreted by Lu et al. as the evidence of a slight repulsion
between the Be atom and the fullerene cage in the Be@C$_{60}$
complex. Our calculations show that the Be's 2$s$ orbital in the
Be@C$_{60}$ complex lies 0.14 eV lower than the C$_{60}$'s HOMO
(see in Figure\ \ref{fig1}). (The Be's 2$s$ orbital in isolated Be
atom lies 0.69 eV lower than the C$_{60}$'s HOMO.) Furthermore,
C$_{60}$'s LUMO-HOMO gap does not change upon the endohedral
doping, and both orbitals are just slightly lowered by 0.01 eV.
Thus, there is little evidence of the ``repulsive interaction'',
according to the definition of Lu at.al., between Be and C$_{60}$
within the Hartree-Fock calculations, even if the calculated
interaction energy is positive.

The Mulliken population analysis of our HF calculations with the
6-31G$^{**}$ molecular basis set gives the charge of +0.03 at Be
compared with -0.14 in the DFT calculation of Lu et.al with the
same basis set. This makes the concept of a slight hybridization
of between Be and the fullerene orbitals unacceptable within the
Hatree-Fock method. However, strictly speaking, the Mulliken
charges are not suitable basis set independent parameters for
analysing the non-chemical interactions.

Then, we consider the local electron density at the beryllium
nucleus. The electron density for a polyatomic system is given by
the standard relation, $\rho({\bf{r}})=\Sigma_n
\kappa_n|\varphi_n({\bf{r}})|^2$, where $\varphi_n({\bf{r}})$ are
normalized molecular orbitals, $\kappa_n$ -- corresponding
occupation numbers. In our case, the orbitals are the Hartree-Fock
orbitals calculated with Danning's cc-pVTZ molecular basis set
(\cite{Dunning-89}) used in a Cartesian form. We employed two
methods for the evaluation of the electron density at the nucleus
which is formally located at the coordinate origin, $r=0$. At
first, we added narrow Gaussian s-functions (with exponents
$\alpha$ between $10^3$ and $10^8$ in inverse squared Bohr radius,
$a_B^{-2}$) to the standard basis set. Then, we extrapolated the
local electron density near the nucleus to the $r=0$ point by
using the known (the consequence of the Kato theorem
\cite{Kato-57}) non-relativistic asymptotic of the electron
density for many-electron system at the Coulomb center with charge
$Z$, $(d\rho(r)/dr)_{r=0}=-2Z\rho(0)$. The coincidence of the both
values of electron density at the Be nucleus with up to 0.02 a.u.
(the atomic unit of density is equal to one electron in cubic Bohr
radius, $a_B^{-3}$), i.e. about 0.05\%, reflects the level of
numerical error for our method.  The calculated electron densities
at single Be atom, metallic Be and Be@C$_{60}$ are summarized in
Table\ \ref{tab2}. (Details of the procedures for calculations of
electron density at metallic Be will be presented elsewhere.)

\begin{table}[h]
\caption{Calculated electron density (in a.u.) at the Be nucleus.}
\label{tab2}
\begin{tabular}{lccccc}
\tableline
   &
    \multicolumn{4}{c}{Orbitals} \\
                   \cline{2-5}
   &
     1-st &
         2-nd &
             Others &
                 Total

                        \\ \hline

Be@C$_{60}$  & 34.22 & 1.24  & 0.02  & 35.48 \\
Be atom      & 34.25 &  1.13 & -  & 35.38\\
Be metal     & 34.11 &  0.32 & 0.33  & 34.78 \\

\tableline
\end{tabular}
\end{table}

Our results show that
$$
\frac{\rho(0)_{\textrm{Be}@\textrm{C}_{60}}-\rho(0)_{\textrm{Be
metal}}}{\rho(0)_{\textrm{Be metal}}}100\% \simeq 2.0\% ,
$$
i.e. a 2\% decrease of the electron density at the nucleus from
Be@C$_{60}$ to metallic Be what is in qualitative agreement with the
experimentally determined change of the decay rate at 5$^{\circ}K$.
Though, the absolute value is somewhat larger then the measured 1.5\%
as well as the value of 1.7\% obtained by the DFT calculations
\cite{Ohtsuki-07,Morisato-08}. On the other hand our value is in
excellent agreement with experimental data of Kraushaar et al.
\cite{Kraushaar-53} of direct measurement of the $^7$Be half-life in
the metal source $T_{1/2}^{^7\textrm{Be metal}} = 53.61\pm 0.17$ days
and the half-life of $^7$Be in $^7$Be@C$_{60}$
$T_{1/2}^{^7\textrm{Be}@\textrm{C}_{60}} = 52.47\pm 0.04$ days from
the work \cite{Ohtsuki-07}:
$$
\frac{T_{1/2}^{^7\textrm{Be metal}} -
T_{1/2}^{^7\textrm{Be}@\textrm{C}_{60}}}{T_{1/2}^{^7\textrm{Be
metal}}}100\% \simeq 2.1\% .
$$

In the present paper we do not discuss these minor quantitative
discrepancies neither between available experimental data, nor
between various theoretical results. Experimental data require the
further specification. As to numerical results it is necessary to
take into account that both methods of calculation contain
approximations. The Hartree-Fock method correctly takes account of
the exchange, but lacks for the electronic correlations. The model
density functionals, on the other hand, make approximations for both
the exchange and the correlations. To make reasoning about the
accuracy of the methods, especially in the case of the
non-chemically, weakly  bounded molecules, one would refer to a more
robust theories like the coupled clusters or the configuration
interaction methods. Instead, we are focussing on a qualitative
phenomenon --- the electron density difference between Be@C$_{60}$
and isolated Be atom. The corresponding relative decrease of the
electron density in our calculations is $0.28$\%, in qualitative
accord with the value of 0.17\% in \cite{Ohtsuki-07,Morisato-08}. The
reason of the enhanced electron density in Be@C$_{60}$ might be a
slight hybridization of Be's and C$_{60}$'s orbitals, as suggested in
\cite{Lu-02}, so that the beryllium grabs the electron density from
the fullerene. In that case, some of fullerene's orbitals would
contribute to the electron density at Be. However, only two orbitals
(those originated from atomic 1$s$ and 2$s$) make apparent
contribution to the electron density at the Be nucleus. Thus, the
hybridization concept has no support from the Hartree-Fock results.

\begin{figure}[h]
\includegraphics[width=10cm]{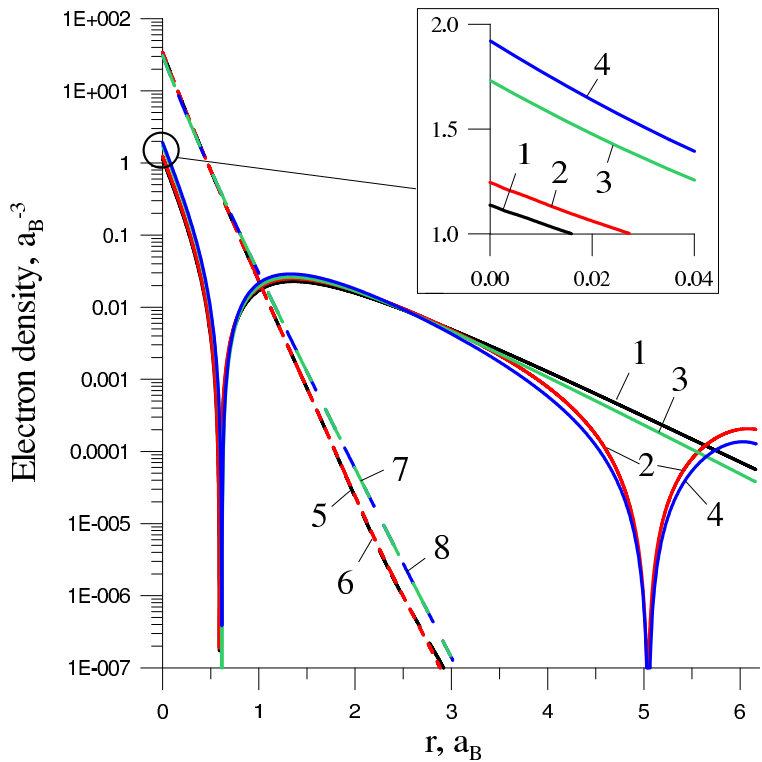}
\caption{Color online. Electron densities of 1$s$ and 2$s$ states for
isolated Be atom, and the first and the second orbitals of the Be
atom in the Be@C$_{60}$ complex: 1 (5) --- 2$s$ (1$s$) Be,
Hartree-Fock; 2 (6) --- 2$s$ (1$s$) Be@C$_{60}$, Hartree-Fock; 3 (7)
--- 2$s$ (1$s$) Be@C$_{60}$, model potential in Figure\ \ref{fig3}; 4
(8) --- 2$s$ (1$s$) Be@C$_{60}$, model potential with model
potential well in Figure\ \ref{fig3}.}
\label{fig2}
\end{figure}

The structure of the Be is therefore changed due to the potential
effect of the fullerene. In particular, fullerene's electrostatic
field might prevent beryllium's electrons to spread out the cage
acting as a strong repulsive potential wall. In that case, a more
compact 1$s$ orbital would not be affected. However, the 2$s$ one
would rapidly vanish after certain distance from the center and
would be compressed (due to the normalization) in the internal
region resulting in the enhanced density at the nucleus. This
concept, which also was expressed in \cite{Ohtsuki-07}, is in full
accord with the data from Table\ \ref{tab2} as well as with the
increased energy of the 2$s$ orbital of Be in C$_{60}$.
Nevertheless, a more detailed analysis of the spherically averaged
electron density curves for the 1$s$ and 2$s$ orbitals of atomic
Be and Be in C$_{60}$ presented on Figure\ \ref{fig2} rules out
the concept of {\it repulsive} potential wall. Indeed, Be's 2$s$
orbital inside the fullerene, though tends to zero at $r \simeq
5a_B$, then rapidly increasing and becomes even larger than the one of
isolated Be atom. Below, we suggest a different solution to this
intriguing issue.


\begin{figure}[h]
\includegraphics[width=8cm]{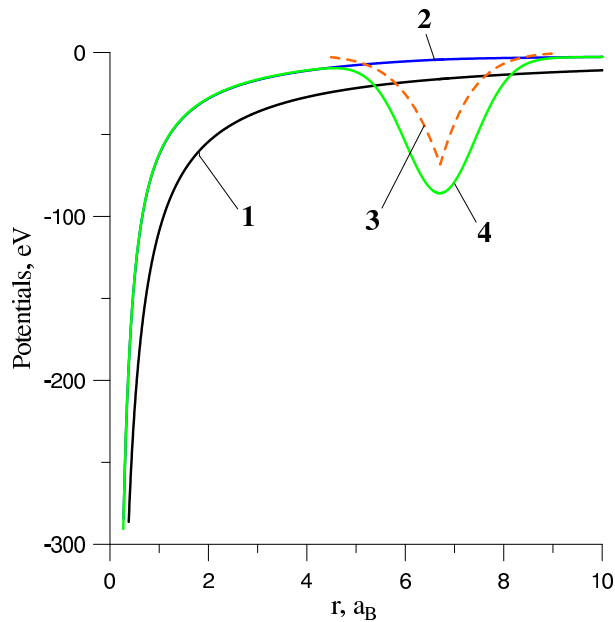}
\caption{Color online. Potentials for the Be 2$s$ orbital in the
Be@C$_{60}$ complex: 1 --- Coulomb potential; 2 --- model potential
(screened Coulomb potential); 3 --- potential well extracted from
Hartree-Fock electron density; 4 --- model potential ``2'' with model
potential well.} \label{fig3}
\end{figure}

The electron density of the 2$s$ orbital vanishes at $r \simeq
5a_B$ since the corresponding wave function crosses zero and
changes its sign. Zero probability for the electrons to occupy the
vicinity of $r \simeq 5a_B$ imitates the repulsive core for the
2$s$ electrons of Be in C$_{60}$. However, the origin of the
additional node in the wave function is the {\it attractive}
potential well at $5a_B \lesssim{} r \lesssim 8a_B$. Spherically
averaged electrostatic potential extracted from the Hartree-Fock
electron density of C$_{60}$ (dashed-line curve in Figure\
\ref{fig3}) is attractive. To illustrate the phenomenon, we
designed a model spherical potential for the 2$s$ orbital of Be
(see in Figure\ \ref{fig3}). It consists of the screened Coulomb
potential as well as the spherical attractive potential well
centered at $r \simeq 6.7a_B$. The screening constant was chosen
to reproduce qualitatively the 2$s$ orbital for the isolated atom.
It turns out, that the addition of the attractive potential into the
model potential results in the appearance of the second node for 2$s$
orbital, in the increase of its  energy
( in full agreement with the result of HF calculation, see in Figure\
\ref{fig1}) and also in the increase of the electron density at $r = 0$.


\section{Analytically solvable model}

\begin{figure}[h]
\includegraphics[width=17cm]{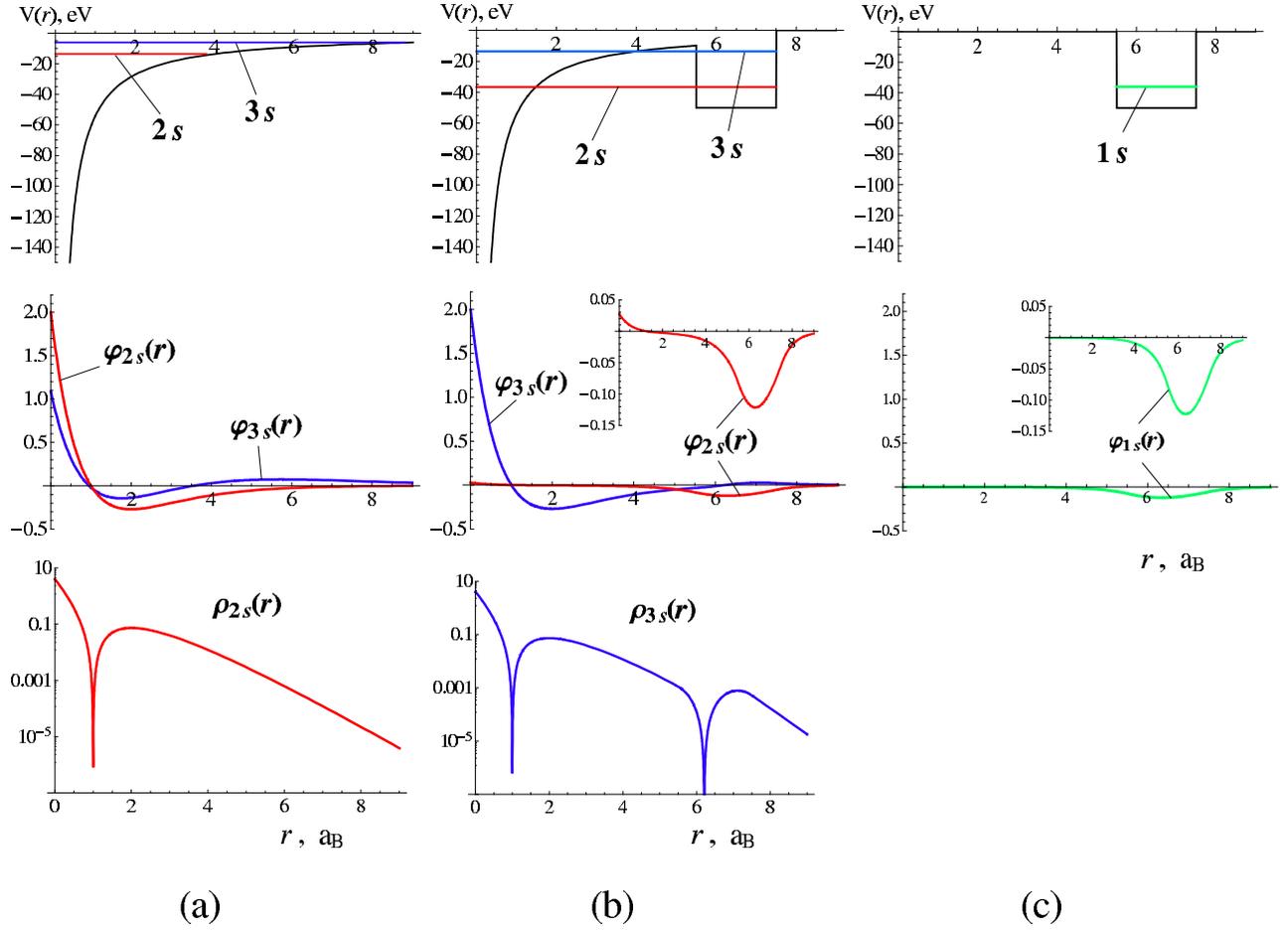}
\caption{Color online. Energy levels, electron wave functions, and
electron densities of 2$s$ and 3$s$ states in: (a) Coulomb potential;
(b) Coulomb potential combined with a spherical potential layer. (c)
Energy level and electron wave function of 1$s$ state in a single
spherical potential layer.} \label{fig4}
\end{figure}

The reason of this phenomenon becomes clear from the following
analytically solvable model. Let us compare the 2$s$ and 3$s$
electron wave functions and the energy levels in the Coulomb potential
(Figure\ \ref{fig4} (a) ) and in the new potential shown in Figure\
\ref{fig4} (b). (This is the same Coulomb potential combined with the
spherical potential layer.)


We consider here only $s$ wave functions $\varphi_{ns}(r)$, because
$\varphi_{np}(0)=0$ for all $p$ states in non relativistic limit, and
these $p$ states do not give a contribution to the decay of $^7$Be.
The electron radial wave functions for the combined potential in
Figure\ \ref{fig4} (b) are

$$
\varphi_{s}(r)  =
\left \{
\begin{array}{lll}
a_1 \exp(-\kappa{}r)_1F_1(1-Z/\kappa;2;2\kappa{}r) , & 0 \leq r < R_1, \\

a_2 (\sin(kr)+b_2\cos(kr))/r , & R_1 \leq r < R_2 , \\
a_3 \exp(-\kappa{}r)/r , & R_2 \leq r .
\label{eq:WF}
\end{array} \right.
$$

Here, $_1F_1$ is the confluent hypergeometric function, the wave
numbers are $\kappa = \sqrt{2m|E|}$, $k = \sqrt{2m(E -V)}$, where $E$
is the energy of the state ($E,V<0$). The coefficients $a_{1-3}$,
$b_2$ and the energy $E$ are determined from the conditions of
continuity and differentiability of the wave function at the
boundaries $r=R_1$ and $r=R_2$, where $R_{1,2}$ are internal and
external radiuses of the potential layer correspondingly.

\begin{figure}[h]
\includegraphics[width=8cm]{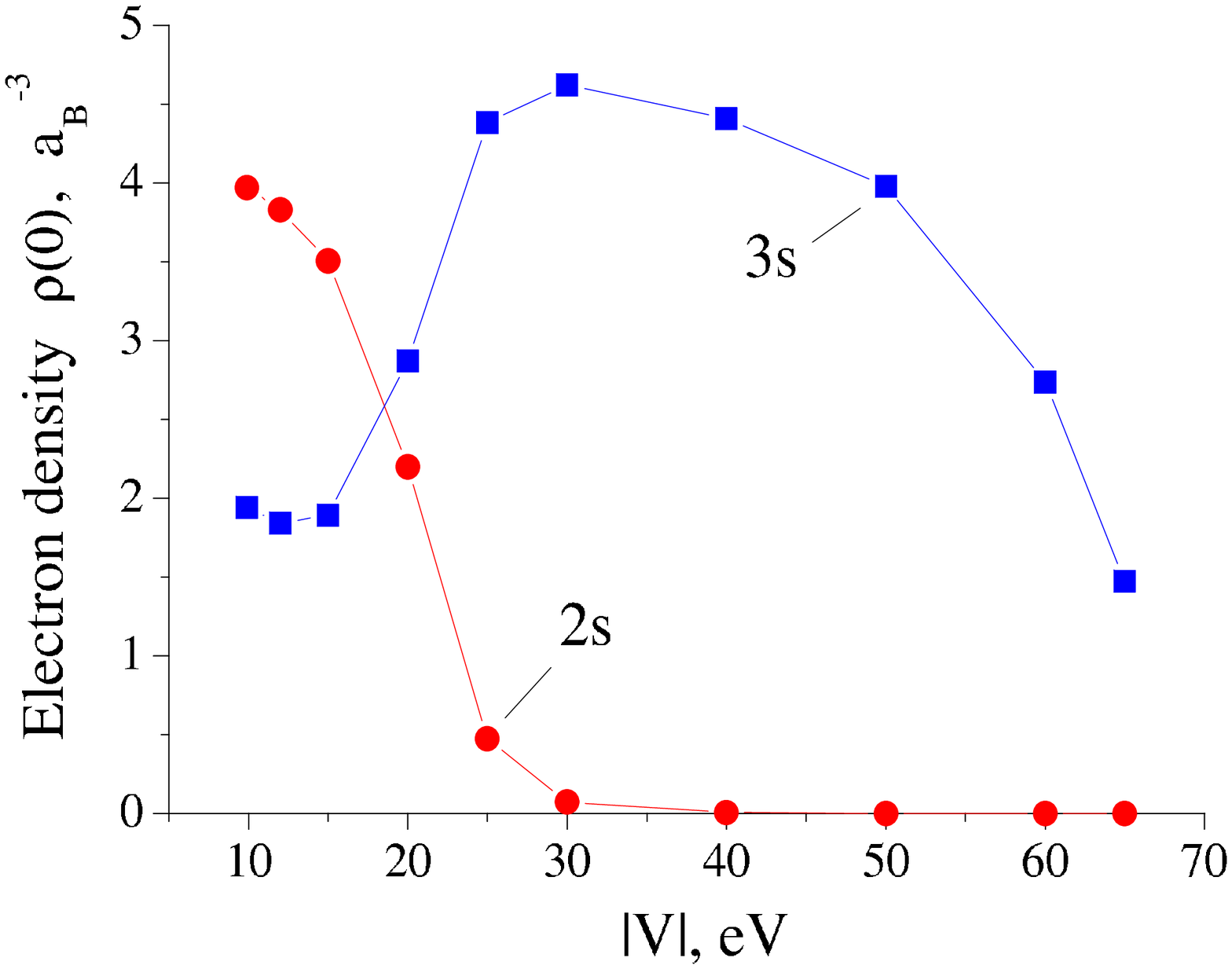}
\caption{Color online. Electron densities of 2$s$ and 3$s$ states at
the point $r=0$ as a function of spherical potential layer depth
$V$.} \label{fig5}
\end{figure}

For definiteness, we take charge $Z=2$ for the Coulomb potential and
depth $V=-50$ eV for the spherical layer located between $R_1 = 5.5$
a$_B$ and $R_2 = 7.5$ a$_B$. The energy and the electron density at
the nucleus of the 2$s$ state in this potential are $E_{2s}=-13.61$
eV and $\rho_{2s}(0)=4$ a$_B^{-3}$, correspondingly. If we add the
spherical potential layer to the Coulomb potential, as it is shown in
Figure\ \ref{fig4} (b), and will gradually increase its depth up to
the value of 50 eV, we will see the following. The energy of the 2$s$
state decreases; its wave function gradually moves from the region of
the Coulomb potential to the region of the spherical layer and
becomes similar to the wave function of the 1$s$ state in the
isolated potential layer shown in Figure\ \ref{fig4} (c). The
electron density of the 2$s$ state at the nucleus decreases according
to Figure\ \ref{fig5}.
%
%
At the same time the 3$s$ state gradually takes up the space region
occupied previously by the 2$s$ state. Depending on the width of the
spherical potential layer and on its depth $V$, the energy of the
3$s$ state in the new combined potential may lie both above and below
of the 2$s$ energy level in the Coulomb potential, and may have
either a higher or a smaller electron density at the nucleus (see in
Figure\ \ref{fig5}).

Qualitatively, the mechanism of the variation of electron density of
the 3$s$ state at the nucleus, shown in Figure\ \ref{fig5}, follows.
The electron density in the origin, $\rho_{ns}(0)$, depends on the
energy of the state $E_{ns}$. For example,
$\rho_{ns}(0)\propto{}|E_{ns}|^{3/2}$ in the Coulomb potential,
$\rho_{ns}(0)\propto{}|E_{ns}|$ in the infinite potential well and so
on. The absolute value of the energy, $|E_{3s}|$, of the 3$s$ state
increases, when the depth $|V|$ of the spherical potential layer
grows. At the same time, gradual redistribution of the 3$s$
electronic density between the area of Coulomb potential and the
spherical potential layer starts. For a relatively small $|V|$, the
space occupied by the 3$s$ wave function in the area of the potential
spherical layer increases comparatively slowly, because the area of
localization of the wave function  practically does not vary, and the
spherical potential layer does not have its own binding state. If the
layer has a large depth, the situation changes. There is a binding
state in the deep isolated spherical potential layer now. The energy
of the 3$s$ state in the joint potential verges towards the energy of
such binding state, moving simultaneously from the position of the
binding state in the pure Coulomb field. A part of the wave function
occupies the forbidden for classical movement area (between the
Coulomb potential and the spherical potential layer), and the
considerable enhancement of the 3$s$ electron density occurs in the
area of the spherical potential layer. The reduction of the wave
function in the area of the Coulomb potential is not compensated
anymore by growth of the $|E_{3s}|$ energy, and the electronic
density at the nucleus decreases fast, returning at first to the
initial value (see in Figure\ \ref{fig5}), and tends to zero at the
further increase of the spherical potential layer depth.

By applying the above considerations to the Be@C$_{60}$ system one
concludes the following. The C$_{60}$ fullerene modifies the Coulomb
potential of the Be atom in such way, that the 3$s$ state in the
joint potential occupies approximately the same position as the 2$s$
state in the isolated Be atom. That is, the 3$s$ state has
approximately the same energy and practically the same electron
density inside the Be atom as the 2$s$ state in Coulomb potential.
Moreover, corresponding wave function $\varphi_{3s}(r)$ has the
second node, which imitates the repulsion of the electrons of the Be
atom from the C$_{60}$ cage at $r \simeq 5a_B$.

As regarding the Be 2$s$ state, the energy of this state becomes
considerably smaller in the new potential. Furthermore, its wave
function moves to the potential well formed by the fullerene.

In that way, the small increase of electron density at the Be nucleus
has casual character. One could obtain another result with different
parameters of the ``fullerene potential well'' (for example, by
doping of C$_{60}$ with certain atoms). Thus, we obtain infrequent
possibility to control the $^7$Be decay by means of a non-chemical
interaction. In the considered particular case, this is the
electrostatic interaction between the Be atom and the fullerene.

To confirm, that the proposed mechanism does not depend on the model,
we also have considered two different analytically solvable models --
a particle inside two ``independent'' and two connected spherical
potential wells. The results obtained are in a good qualitative
agreement with those described above.

\section{Comparison with repulsive core in $\alpha$-$\alpha$ and $N$-$N$ interactions}

It is interesting to note in the end, that the problem considered
here is not physically new and has a vague similarity with the well
known problem of the repulsive core in nuclear physics. The concept
of repulsive core in $\alpha$-$\alpha$ and $N$-$N$ interactions had
been accepted as correct right until the seventies of the last
century. This repulsive core arose at small distances as the
consequence of the Pauli exclusion principle. It was established
later that this concept was simplified. In some cases, the repulsive
core must be understood in terms of the nodal wave function for
relative motion in an attractive potential (see in Refs.
\cite{Saito-69,Neudatchin-83} and references therein).

 In other
words, the zero probability for particle to occupy certain region can
be achieved both by the infinite potential wall (repulsive core) and
by the node of the wave function in an attractive potential. In this
sense the phenomenon considered in the present paper resembles the
above mentioned effects in the $\alpha$-$\alpha$ and the $N$-$N$
interactions.

\section{Conclusion}

In summary, according to our Hartree-Fock calculations with the
electronic correlations accounted at the MP2 level, the lowest
energy singlet configuration of Be atom encapsulated in C$_{60}$
is the one at the center of the fullerene, -- in accordance with
previous DFT studied. The Be atom resides in the attractive region
of the Van-der-Waals interaction with the interaction energy of
about -0.6 eV, --- in contrast with +1.05 eV from the DFT. Thus,
the $^7$Be@C$_{60}$ complex is stable with respect to decay to
$^7$Be and C$_{60}$. The HF electron density at the beryllium
nucleus in Be@C$_{60}$ exceeds the one in metallic Be by 2\% in
qualitative agreement with the relative difference of the
corresponding $^7$Be EC decay rates measured by Ohtsuki et.al. as
well as with the DFT calculations. The electron density at the Be
nucleus in $^7$Be@C$_{60}$ also exceeds the one in isolated Be
atom. The origin of this increasing is neither the modification of
Be's 2$s$ orbital in C$_{60}$ because of the hybridization with
the fullerene orbitals nor the repulsive potential wall at the
region of the fullerene's atoms. Rather the replacement of the Be
2$s$ state by 3$s$ orbital in the new potential, which is the
joint Coulomb potential of Be atom and the attractive effective
potential well generated by the fullerene. The 3$s$ state has
additional node at distance $r \simeq 5a_B$ from the center. This
node imitates the repulsion between electrons of the Be atom and
the C$_{60}$ cage, which has direct physical analogy in the theory
of $\alpha$-$\alpha$ and $N$-$N$ nuclear interactions.

\section{Acknowledgements}

E.~Tkalya thanks Prof. Frederik Scholtz and Dr. Alexander Avdeenkov
for the hospitality and for the given opportunity to do a part of
this work at the National Institute for Theoretical Physics, South
Africa.

\end{document}